\documentclass[adp,a4paper,
]{w-art}
\usepackage{times,cite,w-thm}
\theoremstyle{plain}

\theoremstyle{definition}

\usepackage[]{graphicx}
\begin{document}
\DOIsuffix{theDOIsuffix}
\Volume{16}
\Month{01}
\Year{2007}
\pagespan{1}{}
\Receiveddate{XXXX}
\Reviseddate{XXXX}
\Accepteddate{XXXX}
\Dateposted{XXXX}
\keywords{Relaxation, classical mechanics, ideal gas law.}



\title[Relaxation of ideal classical particles in a one-dimensional box]
{Relaxation of ideal classical particles in a one-dimensional box}

\author[Florian Gebhard]{Florian Gebhard%
\footnote{Corresponding author\quad  
E-mail:~\textsf{\scriptsize florian.gebhard@physik.uni-marburg.de}, Phone: 
+49\,6421\,282\,1318, Fax: +49\,6421\,282\,4511}}
\address{Department of Physics, 
Philipps-Universit\"at Marburg,
D-35032 Marburg, Germany }
\author[Kevin zu M\"unster]{Kevin zu M\"unster}
\begin{abstract}
We study the deterministic dynamics of
non-interacting classical gas particles confined to a one-dimensional box
as a pedagogical toy model 
for the relaxation of the Boltzmann distribution towards equilibrium.
Hard container walls alone
induce a uniform distribution of the gas particles at large times.
For the relaxation of the velocity distribution 
we model the dynamical walls by independent scatterers. The Markov
property guarantees a stationary but not necessarily thermal velocity 
distribution for the gas particles at large times.
We identify the conditions for physical walls where the stationary
velocity distribution is the Maxwell distribution.
For our numerical simulation we represent the wall particles
by independent harmonic oscillators.
The corresponding dynamical map for oscillators with a fixed phase
(Fermi--Ulam accelerator) is chaotic for mesoscopic box dimensions.
\end{abstract}
\maketitle  

\section{Introduction}

Classical mechanics and electrodynamics, quantum mechanics and 
quantum field theory are based on a deterministic time evolution.
For given initial conditions, the particle positions and velocities
or the values of the electromagnetic and Schr\"odinger fields are known for
all times. Therefore, a lecturer of an
introductory course on statistical mechanics 
faces the demanding task to justify a 
probabilistic description of physical processes
on macroscopic time and length scales.

Qualitative arguments which emphasize the importance of the thermodynamic
limit are helpful. 
Students can certainly understand intuitively that information 
gets lost by destructive interference of individual signals:
a man who listens simultaneously to ten thousand discussing philosophers 
receives as much information as if he was herding a flock of sheep.
Nevertheless, it is always helpful to provide explicit calculations
for simple examples where the student can explicitly see that the outcome of
a measurement on a macroscopic deterministic system 
can be derived equivalently from the consideration
of a probabilistic substitute system.

In this work, we provide such a simple example.
We consider a toy model of classical non-interacting point particles
in a box. We assume that our initial conditions factorize in the three
spatial coordinates. Therefore, without a loss of applicability to
three dimensions, we focus on one spatial dimension.
We investigate how an initially 
non-thermal Boltzmann distribution function
relaxes to thermal equilibrium for large times.
As we develop our model, we shall encounter
the known requirements for a
stochastic description of a deterministic system, e.g.,
the coupling of the gas particles to a reservoir.
It is the purpose of this work to show step by step how 
the relaxation of independent point particles is accomplished,
without referring to concepts of statistical mechanics
such as the entropy or the Boltzmann {\sl Sto\ss zahl Ansatz}.
For a thorough discussion of one-dimensional models
which employ the Boltzmann equation, see Ref.~\cite{Ernst}.

In Sect.~\ref{sec:model} we specify the `swimming-lane' model
for the particle propagation and the prediction for the Boltzmann
distribution from statistical mechanics.
In Sect.~\ref{sect:positionrelax} we consider hard walls 
where the distribution of particle velocities remains constant in time. 
We show analytically that the particles are homogeneously distributed
for long times. The relaxation of the particle velocities 
requires dynamical walls where the gas particles exchange energy 
with wall particles. 
In Sect.~\ref{sec:velocity-relax} we consider wall particles whose
velocities are chosen from a probability distribution.
They scatter elastically with the gas particles.
We analytically determine necessary conditions for a `physical wall'
where the gas particles' velocity distribution relaxes
to a Maxwell distribution. 
In Sect.~\ref{sect:numerics}, we investigate numerically 
the deterministic time evolution of the gas particles 
from their elastic scattering against the wall particles 
which we model as harmonic oscillators.
A summary and conclusions, Sect.~\ref{sec:final}, closes our presentation.

\section{Model for classical particles in a box}
\label{sec:model}

One of the simplest model systems for the relaxation to thermal equilibrium
are ideal classical particles in a box. We consider a large number $N\gg 1$ 
of non-interacting 
point-particles $g_{\ell}$ of mass~$m$ which are fully characterized by their
velocities $v_{\ell}$ and their positions $x_{\ell}$ ($|x_{\ell}|\leq a$)
in a one-dimensional box with walls at $|x|=a$. 
Since the particles do not interact with each other 
we can treat them separately. They move independently 
between the walls such as swimmers lap in their individual lanes
(swimming-lane model).

\subsection{Initial conditions for the box particles}

At time $t=0$, we presume that all particles start at position $x=0$
in the middle of the box. Their starting velocities 
are distributed symmetrically,
i.e., for each particle~$g_{\ell}$ with velocity $v_{\ell}$ 
there is a particle~$\overline{g}_{\ell}$ with velocity
$v_{\overline{\ell}}=-v_{\ell}$. 
The probability distribution for the event to find a particle
with velocity~$v$ is denoted by the positive, normalized
function $P_0(v)=P_0(-v)\geq 0$, $\int_{-\infty}^{\infty} {\rm d}v P_0(v)=1$.
The typical velocity of the particles at $t=0$, $v_{\rm typ}$,
obeys $v_{\rm typ}^2=\langle v^2\rangle_0
= \int_{-\infty}^{\infty} {\rm d}v\, v^2 P_0(v)$.
We assume that all moments of the velocity distribution exist,
$\langle v^n\rangle_0<\infty$.

\subsection{Predictions from statistical mechanics for large times}

In our very dilute gas, 
the interaction between the particles and the walls, 
not among the particles themselves, 
is responsible for the thermal equilibrium at large times,
as it is expressed, e.g., in the ideal gas law,
\begin{equation}
p V= N k_{\rm B} T \; .
\label{eq:ideal-gas-law}
\end{equation}
Here, $p$ is the pressure which the $N$~particles exert onto 
the walls of the container of volume $V=2a$;
$k_{\rm B}$ is the Boltzmann constant which permits us
to express the energy content of the wall in terms of the 
familiar temperature~$T$~\cite{Reif}.

For our system of classical particles,
classical statistical mechanics makes elaborate predictions 
beyond~(\ref{eq:ideal-gas-law})
about the particles' properties in thermodynamic equilibrium.
Let $\Delta N(x,v,t)$ be the number of particles which can be found
in the mesoscopic 
position interval $[x,x+\Delta x[$ ($\Delta x = a/\sqrt{N} \ll a$)
with velocities from the mesoscopic interval 
$[v,v+\Delta v[$ ($\Delta v =v_{\rm typ}/\sqrt{N}\ll v_{\rm typ}$)
at time $t$. We write 
\begin{equation}
\Delta N(x,v,t) = N n(x,v,t)(\Delta x)( \Delta v)\; .
\label{eq:introduce-Boltzmann-f}
\end{equation}
In the thermodynamical limit, 
$N\to \infty$ so that $(\Delta x)(\Delta v)\to 0$,
$n(x,v,t)$ approaches the familiar Boltzmann distribution function
$f(x,v,t)$~\cite{kampen}. From the Boltzmann distribution function 
we obtain the distribution functions 
for the particle positions $W(x,t)$ and their velocities $P(v,t)$
as
\begin{equation}
W(x,t)=\int_{-\infty}^{\infty}{\rm d}v f(x,v,t) 
\quad , \quad
P(v,t)=\int_{-a}^{a} {\rm d}x f(x,v,t)  
\; .
\end{equation}
At time $t=0$, we start from
\begin{equation}
f(x,v,t=0)= \delta(x) P_0(v)
\label{eq:Anfangsbedingung}
\end{equation}
so that $W(x,t=0)=\delta(x)$ and $P(v,t=0)=P_0(v)$.

For large times, statistical mechanics predicts that
the Boltzmann distribution approaches its
thermal equilibrium $f_{\rm eq}(x,v)$,
\begin{equation}
f(x,v,t\to\infty)= f_{\rm eq}(x,v)= \frac{1}{2a} 
\Theta\left(1-\frac{|x|}{a}\right)
f_{\rm M}(v)
\quad ; \quad
f_{\rm M}(v) = \sqrt{\frac{m}{2\pi k_{\rm B}T}}
\exp\left(-\frac{mv^2}{2k_{\rm B} T}\right)\; ,
\label{eq:f-equilibrium}
\end{equation}
where $\Theta(x)$ is the Heaviside step-function.
The particles are equally distributed in the box,
$W_{\rm eq}(x)=W(x,t\to\infty)=(1/2a)\Theta(1-|x|/a)$,
and their velocity distribution is given by Maxwell's expression,
$P(v,t\to\infty)=P_{\rm eq}(v)=f_{\rm M}(v)$.

\section{Relaxation of particle positions}
\label{sect:positionrelax}

Thus far, we have not yet specified the properties of the walls.
We start with hard walls so that the particles are perfectly reflected:
Every particle~$g_{\ell}$ with velocity $v_{\ell}$
which is reflected at the right wall
($v_{\ell}>0 \mapsto v'_{\ell}=-v_{\ell}$) has its mirror 
particle~$\overline{g}_{\ell}$ with velocity 
$v_{\overline{\ell}}=-v_{\ell}$ which is reflected
at the left wall 
($v_{\overline{\ell}}\mapsto v'_{\overline{\ell}}=-v_{\overline{\ell}}
=v_{\ell}$). 
Consequently, the velocity distribution of the particles does not change
in time, $P(v,t)=P_0(v)$.

\subsection{Free Boltzmann equation in the absence of walls}

If the walls were absent, all particles start at $x=0$ and propagate
freely according to Newton's first law,
\begin{equation}
x_{\ell}(t)=v_{\ell} t \; .
\label{eq:free-propagation}
\end{equation}
Consequently, the particles at position $x$ with velocity $v$ at time $t$ 
arrive at position $x+v\Delta t$ with velocity $v$ at time $t+\Delta t$.
This free propagation in phase space implies
\begin{equation}
n(x,v,t)=n(x+v\Delta t,v,t+\Delta t) \; .
\end{equation}
In the thermodynamic limit and for small $\Delta t$,
the Taylor expansion of this equation implies the free Boltzmann equation,
\begin{equation}
\frac{\partial f(x,v,t)}{\partial t} 
+ v \frac{\partial f(x,v,t)}{\partial x} =0 \; .
\end{equation}
Its general solution is $f(x,v,t)=g(x-vt,v)$. Together with the 
initial condition~(\ref{eq:Anfangsbedingung}) we find
\begin{equation}
f_{\rm free}(x,v,t)=\delta(x-vt)P_0(v) \; .
\label{eq:free-Boltmann-solution}
\end{equation}
The velocity distribution does not change in time, $P(v,t)=P_0(v)$.
The position distribution broadens in time,
$W(x,t)=P(x/t)/t$. The particles which were initially concentrated
at the origin, disperse into a cloud whose density goes to zero as 
a function of time.

\subsection{Free Boltzmann equation in  the presence of hard walls}

In the presence of hard walls, we consider infinite replicas of
the box on the whole $x$-axis. Box number~$n$ is equivalent
to the interval $[na-a,na+a[$ (extended box scheme).
In the extended box scheme,
the particle propagate freely, see eq.~(\ref{eq:free-propagation}).
In order to determine their true position, we have to fold
the extended box scheme back into the interval for $n=0$
(reduced box scheme). Let us consider $v_{\ell}>0$. After
$m$~wall reflections we find the particles
in the box~$m$. The true particle position
is $x_{\ell}(t)=v_{\ell} t -4na=x_{\ell,{\rm free}}-4na$ for even $m=2n$ 
(particle velocity $v_{\ell}$)
and $x_{\ell}(t)=2a-(v_{\ell} t-4na)=2a+4na-x_{\ell,{\rm free}}$ 
for odd $m=2n+1$ (particle velocity $-v_{\ell}$).
Since the initial velocity distribution is symmetric and does not
change in time, $P(v,t)=P_0(v)$, we obtain 
for the distribution function for all~$v$
\begin{equation}
f(x,v,t) =\Theta\left(1-\frac{|x|}{a}\right)
\left(\sum_{n=-\infty}^{\infty}f_{\rm free}(x+4na,v,t)
+
\sum_{n=-\infty}^{\infty}f_{\rm free}(2a+4na-x,v,t)
\right) \; .
\end{equation}
The free Boltzmann distribution~(\ref{eq:free-Boltmann-solution}) 
is symmetric under the transformation $(-x,-v)\to (x,v)$. Therefore,
we can simplify this expression to~\cite{Song}
\begin{equation}
f(x,v,t) =\Theta\left(1-\frac{|x|}{a}\right)
\sum_{n=-\infty}^{\infty}f_{\rm free}(x+2na,v,t)\; .
\end{equation}
We consider the Fourier series of the Boltzmann distribution
with respect to the position coordinate
and find for the Fourier coefficients ($k=n\pi/a$; $n$: integer)
\begin{eqnarray}
f_k(v,t)&=&\frac{1}{2a} \int_{-a}^a {\rm d}x f(x,v,t)e^{-{\rm i}k x}
\nonumber \\
&=& \frac{1}{2a} \int_{-\infty}^{\infty}{\rm d} x
f_{\rm free}(x,v,t) e^{-{\rm i}k x}
\label{eq:Fouriercoefficients} \\
&=& \frac{1}{2a} e^{-{\rm i}k vt }P_0(v)
\nonumber \; .
\end{eqnarray}
The Boltzmann distribution function in the presence of hard walls
and the free Boltzmann distribution function have the same
Fourier coefficients at $k=n\pi/a$ so that 
\begin{equation}
f(x,v,t)=\frac{1}{2a}\Theta\left(1-\frac{|x|}{a}\right)
\sum_{n=-\infty}^{\infty}e^{{\rm i}n(\pi/a) (x-vt)}P_0(v)
\; .
\end{equation}
With this result, we can calculate the probability
distribution for the particle positions. We use the Fourier transformation
for the initial velocity distribution,
\begin{equation}
P_0(v)=\int_{-\infty}^{\infty} \frac{{\rm d}\eta}{2\pi} e^{{\rm i}v\eta}
\widetilde{P}_0(\eta) \quad ,
\quad 
\widetilde{P}_0(\eta) =
\int_{-\infty}^{\infty} {\rm d}v e^{-{\rm i}v\eta} P_0(v)
\label{eq:defFTforP}
\end{equation}
with $\widetilde{P}_0(\eta)= \widetilde{P}_0(-\eta)$ and find
\begin{eqnarray}
W(x,t)&=& \frac{1}{2a} \Theta\left(1-\frac{|x|}{a}\right)
\sum_{n=-\infty}^{\infty}
e^{{\rm i} x n\pi/a}\int_{-\infty}^{\infty}{\rm d} v e^{-{\rm i}ntv\pi/a}P_0(v)
\nonumber \\
&=& \frac{1}{2a}\Theta\left(1-\frac{|x|}{a}\right)
\sum_{n=-\infty}^{\infty}
e^{{\rm i}nx\pi/a}\widetilde{P}_0(n\pi t/a) \; .
\label{eq:final-Boltzmann-hard-walls}
\end{eqnarray}
For each $n\neq 0$, the argument of $\widetilde{P}_0(\eta)$ becomes large
for $t\to\infty$.
We demand
\begin{equation}
\widetilde{P}_0(\eta\to\infty)=0 \;.
\end{equation}
This implies that the initial velocity distribution
cannot contain a macroscopic
number of particles with the same velocity.
Then, for large times,
the sum in~(\ref{eq:final-Boltzmann-hard-walls}) reduces to its 
contribution from $n=0$,
\begin{equation}
W(x,t\to\infty)= \frac{1}{2a}\Theta\left(1-\frac{|x|}{a}\right)  \; .
\end{equation}
For long times, the particles become equally distributed 
in the box with density $\rho_0=N/V=N/(2a)$, 
as expected from thermodynamic considerations.

It is important to note that we know the particle positions and their
velocities exactly at every instant in time. 
Despite this fact, the particle distribution becomes homogeneous 
for long times and `relaxes' to its equilibrium distribution.
For the particle positions this does not come as a surprise because
there is no conservation law which prevents them to 
distribute equally in space.
For the velocity distribution, however, momentum and energy conservation
guarantee that it is independent in time for hard walls
and our symmetric initial distribution. 
The relaxation of the velocity distribution is more involved,
as we discuss in Sec.~\ref{sec:velocity-relax}.

\subsection{Pressure on the hard walls}

The reflection of the particles at the right wall implies
a momentum transfer $\Delta P$ to it. All the particles with velocity $v>0$
which reach the wall within the time span $\Delta t$ transfer the momentum
$2m v$~\cite{Reif}. Therefore,
\begin{equation}
\Delta P=N \int_{0}^{\infty} {\rm d}v (2mv) 
\int_{a-v\Delta t}^{a}{\rm d}x n(x,v,t)
\; .
\end{equation}
For small $\Delta t$ and in the thermodynamic limit,
we find for the pressure in a one-dimensional box,
$p(t)=\lim_{\Delta t\to 0}(\Delta P)/(\Delta t)$, and with $V=2a$
\begin{eqnarray}
p(t) V &=&  2N \int_{-\infty}^{\infty}{\rm d}v \frac{mv^2}{2} (2a) f(a,v,t) 
= 2 N \left(\overline{E} +H(t)\right)\; , \nonumber  \\
\overline{E} &=& \int_{-\infty}^{\infty}{\rm d}v \frac{mv^2}{2} P_0(v)
\label{eq:ideal-gas-law-derived} \; ,\\
H(t)&=&
\sum_{n=1}^{\infty} (-1)^n \int_{-\infty}^{\infty}{\rm d}v (mv^2)
e^{-in\pi vt/a} P_0(v) \; .
\label{eq:pressure-result-hardwall}
\end{eqnarray}
The time-dependent terms, $H(t)$, vanish for $t\to\infty$ because the
Fourier transformed velocity distribution $\widetilde{P}_0(\eta)$ and
its derivatives vanish for large arguments.
We find for $t>0$
\begin{equation}
H(t)=m \sum_{n=1}^{\infty}(-1)^{n+1}
\widetilde{P}_0^{\prime\prime}\left(\frac{nt\pi}{a}\right)
\; ,
\end{equation}
where $\widetilde{P}_0^{\prime\prime}(\eta)$ denotes the second
derivative.

The time-independent term provides the ideal-gas law~(\ref{eq:ideal-gas-law})
in one dimension~\cite{Reif},
$p(t\to\infty)V=2N \overline{E}$ where
$\overline{E}$ is the average energy of a particle.
For the Maxwell distribution, see eq.~(\ref{eq:f-equilibrium}), we find 
$\overline{E}=k_{\rm B}T/2$ so that $p(t\to\infty)V=Nk_{\rm B}T$,
eq.~(\ref{eq:ideal-gas-law}), results.

\subsection{Examples}

As examples, we consider an exponential velocity distribution
and a step-like velocity distribution,
\begin{eqnarray}
P_{\rm ex}(v) &=& \frac{1}{2 v_0}e^{-|v/v_0|} 
\quad , \quad 
\widetilde{P}_{\rm ex}(\eta)=\frac{1}{1+\eta^2v_0^2} \; ,
\label{eq:FTexp}
\\
P_{\rm step}(v)&=& \frac{1}{2 v_0} \Theta\left(1-\frac{|v|}{v_0}\right)
\quad , \quad
\widetilde{P}_{\rm step}(\eta)=\frac{\sin(v_0\eta)}{v_0\eta} \; .
\label{eq:FTstep}
\end{eqnarray}
The particle density at the wall is 
obtained from~(\ref{eq:final-Boltzmann-hard-walls})
\begin{equation}
W(a,t)=\frac{1}{2a} \left(1+2\sum_{n=1}^{\infty}(-1)^n \widetilde{P}_0(n\pi t/a)
\right) \; .
\end{equation}
With the help of Eq.~(27.8.6) in Ref.~\cite{Abramovitz}
and {\sl Mathematica}~\cite{Mathematica} we find for $t>0$
($t_{\rm tr}=a/v_0$)
\begin{eqnarray}
2a W_{\rm ex}(a,t)&=& 
1+2\sum_{n=1}^{\infty}\frac{(-1)^n }{1+(\pi n t/t_{\rm tr})^2}
= \frac{1}{(t/t_{\rm tr})\sinh(t_{\rm tr}/t)} \; ,\\
2a W_{\rm step}(a,t)&=& 
1+ 2 \sum_{n=1}^{\infty} (-1)^n \frac{\sin(\pi n t/t_{\rm tr})}{\pi n t/t_{\rm tr}}
= 1- \frac{s(t/t_{\rm tr})}{t/t_{\rm tr}} \; .
\end{eqnarray}
Here, $s(x)=x-2n$ for $x\in[-1+2n,1+2n]$ so that $|s(x)|\leq 1$.
The transient time $t_{\rm tr}$ is given by the time a typical particle
needs to travel through the system, $t_{\rm tr}=a/v_0$.
After this time, the density at the wall begins to approach
its equilibrium value, $N W(a,t\to\infty)=\rho_0$.
For the exponential distribution, deviations are of order $(t_{\rm tr}/t)^2$
for large times. For the step-like velocity distribution,
we observe saw-tooth oscillations which die out for long times
only proportional to $(t_{\rm tr}/t)$.
This shows that the particles swap back and forth through the box and 
reach their homogeneous distribution only fairly slowly.

\begin{figure}
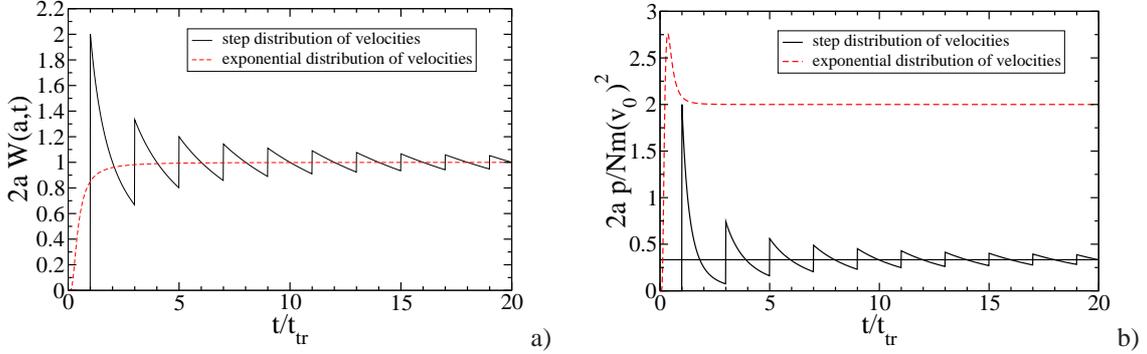

\vspace{12pt}
\includegraphics[width=68mm]{walldensity.eps}~a)
\hfill
\includegraphics[width=68mm]{pressure.eps}~b)
\caption{Time evolution of the particle density at the wall, \textbf{a},
and of the pressure, \textbf{b}, for the step-like and exponential 
velocity distributions. 
Note that the average kinetic 
energy differs for the two
distributions, $2\overline{E}_{\rm step}/(mv_0^2)=1/3$
and $2\overline{E}_{\rm exp}/(mv_0^2)=2$.\label{fig1}}
\end{figure}

The same analysis is readily carried out for the pressure.
With the help of Eq.~(27.8.6) in Ref.~\cite{Abramovitz}
and {\sl Mathematica}~\cite{Mathematica} we find for $t>0$
($t_{\rm tr}=a/v_0$)
\begin{eqnarray}
H_{\rm ex}(t)&=&2 m v_0^2 \sum_{n=1}^{\infty} (-1)^n 
\frac{1-3 (n\pi t/t_{\rm tr})^2}{[1+(n \pi t/t_{\rm tr})^2]^3} 
= \frac{mv_0^2}{2}\left[
-2 + \frac{3+\cosh(2t_{\rm tr}/t)}{2[(t/t_{\rm tr})\sinh(t_{\rm tr}/t]^3}
\right]\; ,\\
H_{\rm step}(t) &=& m v_0^2 \sum_{n=1}^{\infty} (-1)^n
\left[\frac{2\cos(n\pi t/t_{\rm tr})}{(n\pi t/t_{\rm tr})^2} +
\frac{\sin(n\pi t/t_{\rm tr})}{n\pi t/t_{\rm tr}}
\left(1-\frac{2}{(n\pi t/t_{\rm tr})^2}\right)
\right] \nonumber\\
&=&  
\frac{m v_0^2}{2} \left[
-\frac{s(t/t_{\rm tr})}{t/t_{\rm tr}}
+ \frac{3(s(t/t_{\rm tr}))^2-1}{3(t/t_{\rm tr})^2}
-
\frac{s(t/t_{\rm tr})(s(t/t_{\rm tr})^2-1)}{3(t/t_{\rm tr})^3}
\right]\; .
\end{eqnarray}
For large times, $t\gg t_{\rm tr}$, we find for the smooth 
exponential velocity distribution that
$H_{\rm ex}(t\gg t_{\rm tr})\sim (t_{\rm tr}/t)^4$, i.e.,
the pressure corrections vanish quickly as a function of time.
For the step-distribution we observe again a slow decay with
saw-tooth oscillations, 
$H_{\rm step}(t\gg t_{\rm tr})\sim (t/t_{\rm tr})^{-1}$.
The density of particles at the wall and the pressure 
are shown in Fig.~\ref{fig1}.

\section{Relaxation of particle velocities}
\label{sec:velocity-relax}

In order to relax the energy of the gas particles, we introduce
scattering processes in which the gas particles exchange
energy with the walls. 

\subsection{Dynamic wall and scattering events}

In order to simplify the analysis, we assume that the walls are perfectly 
symmetric. Each scattering event which occurs for the gas particle~$g_{\ell}$
with velocity $v_{\ell}$ at the right wall ($x=a$) has a mirror event 
for the particle $\overline{g}_{\ell}$ with velocity $v_{\overline{\ell}}=
-v_{\ell}$ at the left wall ($x=-a$). In this way, our velocity distribution 
remains symmetric around $v=0$ at all times.
In this section we are primarily interested in the relaxation of the
velocity distribution of the gas particles. Therefore, we
focus on the velocity, not on the position of the gas particles.

The right wall is made of $L\gg N\gg 1$~particles which move freely 
in the interval $[a,a+A]$ ($a\gg A>0$). The wall particles 
are scattered elastically at the interval boundaries.
The boundary at $x=a$ is transparent for the gas particles.
The velocities of the wall particles 
are taken from a probability distribution $D(u)=D(-u)$
which obeys the same restrictions as $P_0(v)$, i.e.,
all of its moments exist and its Fourier transform, $\widetilde{D}(\eta)$,
decays to zero for large values, 
$\lim_{\eta\to\infty}\widetilde{D}(\eta)=0$.

The gas particle~$g_{\ell}$ ($\ell=1,\ldots N$) 
which starts with velocity $v_{\ell,0}>0$ at time $t=0$ undergoes
a first scattering event at time $t_1>0$
with the wall particle $w_{\ell,1}$ with velocity $u_{\ell,1}$.
After the scattering process is completed, the gas particle has the velocity
$v_{\ell,1}<0$. At time $t_2>t_1$ it is scattered off the left wall 
by the wall particle $w_{\ell,2}$ whose velocity $u_{\ell,2}$ is independent 
of the velocity of the wall particle $w_{\ell,1}$. 
The same applies to the subsequent scattering processes:
The scattering partners of the gas particles
are all different, $w_{\ell,n}=w_{\ell',n'}$ only if 
$\ell=\ell'$ and $n=n'$.
In this way we assure that the wall has no memory:
the outcome of a scattering event does not influence the 
distribution $D(u)$. 
This assumption expresses the fact that the wall
particles form a `bath' for the gas particles.

Let the wall particles have the mass~$M>m$.
Then, a scattering event between a gas particle with velocity~$v$
and a wall particle with velocity~$u$ results in
\begin{eqnarray}
v'&=& \beta u -\alpha v\; , \nonumber \\
u'&=&r \beta v + \alpha u\; ,\\
r=\frac{m}{M}\quad,\quad \alpha&=&\frac{1-r}{1+r} \quad,\quad 
\beta = \frac{2}{1+r} \nonumber 
\end{eqnarray}
with $0<r<1$, $0<\alpha<1$, and $1<\beta<2$.

\subsection{Master equation for the velocity distribution}

When we focus on scattering events at the right wall, we
must distinguish three cases.
We assume that the wall particle and the gas particle meet at
the left wall boundary, $x=a$.
\begin{itemize}
\item[--]{$u<\alpha v/\beta$}: 

The gas particle is reflected, 
$v'=\beta u-\alpha v<0$.
\item[--] {$\alpha v/\beta\leq u<v$}: 

After the first collision,
the gas particle penetrates the wall
because its velocity is still positive, $u'>v'>0$.
The wall particle is reflected at the wall boundary at $x=a+A$,
and collides with the gas particle again.
The gas particle then has the velocity 
\begin{equation}
v''=\gamma v -2\alpha\beta u<0\quad , \quad 
\gamma=\alpha^2-r \beta^2 \; ,
\end{equation}
and escapes from the wall region because $v''<u''$.
\item[--] {$v\leq u$}:

In this case, the wall particle escapes
from the gas particle. The gas particle penetrates the wall
and collides with the wall particle which was reflected 
at the wall boundary at $x=a+A$.
After the collision, the gas particle has the 
velocity $v'=-\beta u -\alpha v<0$.
It cannot be reached by the wall particle again, $v'<u'$.
\end{itemize}
If the average kinetic energy of the gas particles
and the wall particles is comparable and in the limit $m/M\to 0$, 
the typical wall particle velocities are small compared to 
the typical gas particle velocities. Then, only the first case 
is relevant.

After all gas particles have been scattered $n$~times,
the probability distribution after the $n$th scattering
process ($n=1,2,\ldots$) is given by
\begin{eqnarray}
P(n;v'>0) &=& 
\int_{0}^{\infty}{\rm d} v\, \frac{P(n-1;v)}{\beta}
D\left(\frac{\alpha v-v'}{\beta} \right)
+ 
\int_{0}^{v'/(\alpha+\beta)}{\rm d} v\, \frac{P(n-1;v)}{\beta}
D\left(\frac{v'-\alpha v}{\beta} \right) \nonumber \\
&& +
\int_{v'/(\alpha+\beta)}^{v'}{\rm d} v\, 
\frac{P(n-1;v)}{2\alpha\beta}
D\left(\frac{v'+\gamma v}{2\alpha\beta} \right) 
\; ,
\label{eq:help}
\end{eqnarray}
where we used the fact that our probability distributions are symmetric
around the origin, $P(n;v)=P(n;-v)$, $D(-u)=D(u)$.
We can rewrite this equation in the form
\begin{equation}
P(n;v)= \int_{-\infty}^{\infty}{\rm d}v'\, A(v,v')P(n-1;v')
\label{eq:almost-master}
\end{equation}
where the integral kernel $A(v,v')$ is readily obtained from~(\ref{eq:help}).
Note that the wall has no memory. Therefore, the 
$n$th scattering distribution $P(n;v)$ depends only on
the distribution before the $n$th event, $P(n-1;v)$.
This is the Markov property~\cite{kampen}.
The probability distributions are normalized to unity.
The condition $\int_{-\infty}^{\infty}{\rm d}v\, A(v',v)=1$ implies
that no gas particle is lost which guarantees that 
the probability distribution remains normalized to unity,
$\int_{-\infty}^{\infty} {\rm d}v\, P(n;v)=1$.

In order to elucidate the physical meaning of Eq.~(\ref{eq:almost-master}),
we introduce the artificial time $\tau=n \Delta \tau$ ($\Delta \tau=1$), 
and consider the limit
of a large number of scattering events, $n \gg 1$, so that 
$\Delta\tau \ll \tau$. We write
\begin{equation}
P(n+1;v)-P(n;v)= \int_{-\infty}^{\infty}{\rm d}v'\, 
\left(A(v,v')P(n;v')-A(v',v)P(n;v)\right) \; .
\label{eq:almost-master-2}
\end{equation}
In the limit $n\to\infty$, the time $\tau$ becomes a continuous mesoscopic
time scale and
we obtain a Master equation~\cite{kampen} 
for the velocity distribution $P(\tau;v)$,
\begin{equation}
\frac{\partial P(\tau;v)}{\partial \tau}= 
\int_{-\infty}^{\infty}{\rm d}v'\, 
\left(A(v,v')P(\tau;v')-A(v',v)P(\tau;v)\right) \; .
\label{eq:master}
\end{equation}
This is a gain-loss equation for the number of particles $N(v)=NP(v)$
with velocity $v$. As has been shown in the literature~\cite{kampen},
the master equation~(\ref{eq:master}) has a unique stationary
solution $P_{\rm stat}(v)=\lim_{\tau\to\infty}P(\tau;v)$.

\begin{figure}[ht]
\sidecaption
\includegraphics[width=.5\textwidth]{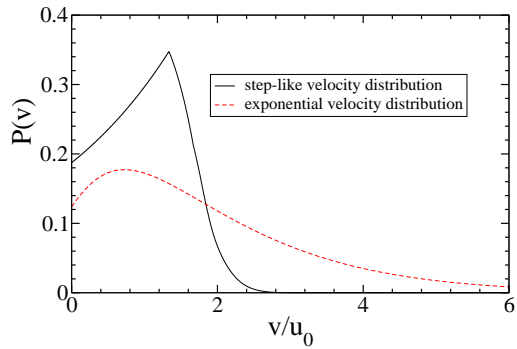}
\caption{Stationary velocity distribution for the gas particles 
after an infinite number of collisions with the wall. The mass ratio
between gas particles and wall particles is $m/M=1/2$.
The velocity distribution of the wall particles is exponential and step-like,
see Eq.~(\protect\ref{eq:walldistributions}).}
\label{fig2}
\end{figure}

In general, the stationary solution of the Master equation~(\ref{eq:master})
is {\em not\/} the Maxwell velocity distribution.
As an example, we show the result for $m/M=1/2$ for
exponential and step-like velocity distributions of the wall particles
in Fig.~\ref{fig2},
\begin{equation}
D_{\rm ex}(u)=\frac{1}{2u_0} e^{-|u/u_0|} \quad, \quad 
D_{\rm step}(u) =\frac{1}{2u_0} 
\Theta\left(1-\frac{|u|}{u_0}\right) \; .
\label{eq:walldistributions}
\end{equation}
The stationary velocity distributions depend
only on the choice of the velocity distributions for the wall particles,
$D(u)$, and not on the initial velocity distribution of the gas particles,
$P_0(v)$. The result is independent of the initial conditions.
Note that a Gaussian velocity distribution
$D(u)$ does {\sl not\/} lead to a Maxwell velocity distribution $P(v)$.

The wall in the above example is not a `physical' wall.
According to the definition given in van Kampen's book~\cite{kampen},
a physical wall relaxes the gas particles' velocities
to a Maxwell velocity distribution.

\subsection{Relaxation to a Maxwell distribution for a physical wall}
\label{subsec:physwall}

For a thermal reservoir, 
its properties must remain essentially unchanged by the interaction
with the gas. If a wall particle transferred or received 
a substantial amount of energy in an individual scattering process
it would carry information about the gas particles. 
Therefore, for a physical wall, we have to demand that the energy transfer
in each collision is small compared to the typical kinetic energy of a 
wall particle. 
This can be guaranteed in the limit $m/M\to 0$.
More specifically, we presume that
the typical kinetic energy of a
gas particle and of a wall particle are of the same order of magnitude
$mv_{\rm typ}^2\approx Mu_{\rm typ}^2$. Then, the typical
velocity of a wall particle is small
compared to the typical velocity of a gas particle, i.e.,
$u_{\rm typ}$ is of the order $\sqrt{m/M}v_{\rm typ}\ll v_{\rm typ}$.

In order to perform the limit $m/M\to 0$, we rewrite 
Eq.~(\ref{eq:help}) in the form
\begin{eqnarray}
P(v'>0) &=& \int_{-v'\beta}^{\infty}{\rm d}x \frac{D(x)}{\alpha}
P\left(\frac{\beta x+ v'}{\alpha}\right)
+ 
\int_{v'/(\alpha+\beta)}^{v'/\beta}{\rm d}x \frac{D(x)}{\alpha}
P\left(\frac{v'-\beta x}{\alpha}\right)
\nonumber \\
&& +
\int_{v'/(\alpha+\beta)}^{v'\alpha/\beta}
{\rm d}x \frac{D(x)}{|\gamma|}
P\left(\frac{2\alpha\beta x-v'}{\gamma}\right) \; ,
\label{eq:transformed-eq}
\end{eqnarray}
where we used $\lim_{n\to\infty}P(n;v)=P(v)$.
Now we are in the position to let $m/M\to 0$ where
$\alpha \approx 1-2m/M \to 1$, $\beta\approx 2(1-m/M)\to 2$, 
$\gamma \approx (1-8m/M)\to 1$.
$D(x)$ is finite only in a region which is small compared to the typical
values for $v'$. Therefore, we may safely extend the 
first integral over the whole real axis, and we may ignore the
contributions from the second and third integral because
$D(v')=0$ for typical values of $v'$. The excluded region, 
$v'\in [0,(m/M) u_{\rm typ}]$, becomes vanishingly small in the limit
$m/M\to 0$.

The resulting integral equation can be simplified by Fourier transformation.
We find, see eq.~(\ref{eq:defFTforP}),
\begin{equation}
\widetilde{P}(\eta) = \widetilde{D}(\beta\eta)\widetilde{P}(\alpha\eta) \; .
\end{equation}
For the normalized, differentiable, and symmetric $\widetilde{P}(\eta)$ 
and $\widetilde{D}(\eta)$, 
the Taylor expansion of this equation 
gives ($\widetilde{D}(0)=1$,  $\widetilde{D}'(0)=0$, $\beta=2$,
$\alpha=1-2m/M$)  
\begin{equation}
\widetilde{P}(\eta) = 
\left( \widetilde{P}(\eta)
-\frac{2m\eta}{M}\widetilde{P}'(\eta)
\right) + 2\eta^2\widetilde{D}''(0) \widetilde{P}(\eta)
\; .
\label{eq:expansion-for-Maxwell}
\end{equation}
Here, we made use of the fact that $\widetilde{D}(\eta)$ is
almost constant in the region where $\widetilde{P}(\eta)$ is finite.
This reflects our assumption that $D(v)$ is a narrow distribution
in comparison to $P(v)$.
The resulting differential equation is readily integrated.
We use 
\begin{equation}
\widetilde{D}''(0)=\int_{\-\infty}^{\infty}{\rm d} uD(u) (-{\rm i}u)^2
\equiv -\langle u^2\rangle \; ,
\end{equation}
and {\em define\/} the wall temperature~$T$ via
\begin{equation}
\frac{M}{2} \langle u^2\rangle = \frac{k_{\rm B}T}{2}
\; ,
\label{eq:equipart}
\end{equation}
which corresponds to the equipartition theorem in one dimension.
Note, however, that in our case $k_{\rm B} T/2$ simply is an abbreviation
for the average kinetic energy of the wall particles.
Then, we find from~(\ref{eq:expansion-for-Maxwell})
\begin{equation}
\widetilde{P}(\eta)= e^{-\eta^2 k_{\rm B}T/(2m)}
\end{equation}
so that we finally obtain the Maxwell velocity distribution for the
gas particles,
\begin{equation}
P(v)= f_{\rm M}(v)= \sqrt{\frac{m}{2\pi k_{\rm B}T}}
\exp\left(-\frac{mv^2}{2k_{\rm B} T}\right) \; .
\label{eq:Maxwell}
\end{equation}
The average kinetic energies of the gas particles and the wall particles
are the same, $M\langle u^2\rangle=m\langle v^2\rangle=k_{\rm B}T$, 
but their typical velocities differ by a factor $\sqrt{m/M}$,
$u_{\rm typ}=\sqrt{\langle u^2\rangle}=
\sqrt{m/M} v_{\rm typ}$ with 
$v_{\rm typ}=\sqrt{\langle v^2\rangle}$.

\section{Numerical simulations for a deterministic system}
\label{sect:numerics}

Finally, we present the results for a numerical simulation
of gas particles in a box with 
a deterministic, physical wall. For simplicity we choose the left wall 
to be perfectly reflecting.

\subsection{Oscillator model}

We represent the right wall by harmonic oscillators of mass~$M$
with frequency $\omega_0$.
Each gas particle $g_{\ell}$ ($\ell=1,2\ldots,N$)
undergoes a sequence of collisions with the wall particles
$w_{\ell,i}$ ($i=1,2,\ldots,L_{\ell}$; $L_{\ell}\gg 1$)
whose free motion is given by 
\begin{equation}
x_{\ell,i}(t)=a - A \cos(\omega_0 t+\phi_{\ell,i}) \quad , \quad 
u_{\ell,i}(t)= (A\omega_0)\sin(\omega_0 t+\phi_{\ell,i}) \; .
\end{equation}
The length~$A$ parameterizes the `penetration depth' of the gas particle
into the wall. We assume it to be a small fraction of the
interatomic distance between the wall's atoms and, therefore, we work with
$A=1\cdot 10^{-12}\, {\rm m}=0.01$\AA.
We choose $a=0.2\, {\rm m}$ for the diameter of the box,
$a/A=2\cdot 10^{11}$.
The typical phonon frequencies in a solid are 
$\omega_0=1.3\cdot 10^{13}\, {\rm Hz}$ 
so that $u_0= A\omega_0=13\, {\rm m/s}$.
The phases $\phi_{\ell,i}\in [0,2\pi[$ 
are chosen at random. This reflects the fact that the motion of the
wall particles is uncorrelated.
We shall comment on the case of equal phases in Sect.~\ref{sect:equalphase}.

For long enough times, we expect that we recover
the situation of Sect.~\ref{subsec:physwall}.
The velocity distribution of the 
wall particles is given by ($T_0=2\pi/\omega_0$)
\begin{equation}
D(u) = \frac{1}{T_0} \int_{0}^{T_0} {\rm d} t
\delta\left(u-u_0\sin(\omega_0 t+\phi)\right) = \frac{1}{\pi} 
\frac{1}{\sqrt{u_0^2-u^2}} \quad \hbox{for}  \quad 
|u|\leq u_0 \; .
\label{eq:oscimodel}
\end{equation}
The Fourier transformed velocity distribution is given by 
a Bessel function $\widetilde{D}(\eta)=J_0(u_0\eta)$ which decays
to zero for large arguments,
$\widetilde{D}(\eta\gg 1/u_0)
\propto 1/\sqrt{\eta u_0}$. 

{}From~(\ref{eq:equipart}) it follows that
the wall temperature is given by $k_{\rm B}T = M u_0^2/2$
because $\langle u^2\rangle=u_0^2/2$ for the velocity 
distribution~(\ref{eq:oscimodel}).
Therefore, at room temperature we find $M\approx 29 \cdot 10^3\, {\rm u}$ 
(${\rm u}$: atomic mass unit), 
i.e., of the order of ${\cal O}(10^2)$~atoms 
in the wall constitute a `wall particle'.
As our mass ratio we set $m/M=10^{-3}$, i.e.,
the gas particles have the mass $m=29\, {\rm u}$ (nitrogen molecule, N$_2$).

We expect that the stationary velocity distribution
of our gas particles becomes a Gaussian distribution~(\ref{eq:Maxwell})
with variance $\langle v^2\rangle=k_{\rm B}T/m=M/(2m)u_0^2=500u_0^2$.
We thus expect $v_{\rm typ}=\sqrt{\langle v^2\rangle}= 23 u_0
\approx 3\cdot 10^2\, {\rm m/s}$ which agrees
with the typical velocity of gas particles in a container at room temperature.
The time for a typical particle to perform a closed loop
through the box is 
$t_{\rm typ}=4a/v_{\rm typ}\approx 1.3 \cdot 10^{-3}\, {\rm s}$.
A good estimate of the number of scatterings 
$C(t)$
which a gas particle has
undergone after the time~$t$ is given by
$C(t)={\cal O}\bigl(t/(10 t_{\rm tr})\bigr)$.

At time $t=0$, we distribute our $N/2=1\cdot 10^7$
gas particles equidistantly in the velocity interval $]0,v_0]$,
$P_0(v\geq 0)=\Theta(1-v/v_0)/(2v_0)$, with $v_0=10u_0=130\, {\rm m/s}$. 
The results for negative velocities are obtained by a reflection
at the axis $v=0$.
Initially, all gas particles start at $x=0$,
$f(x,v,t_0=0)=P_0(v)\delta(x)$.

\begin{figure}
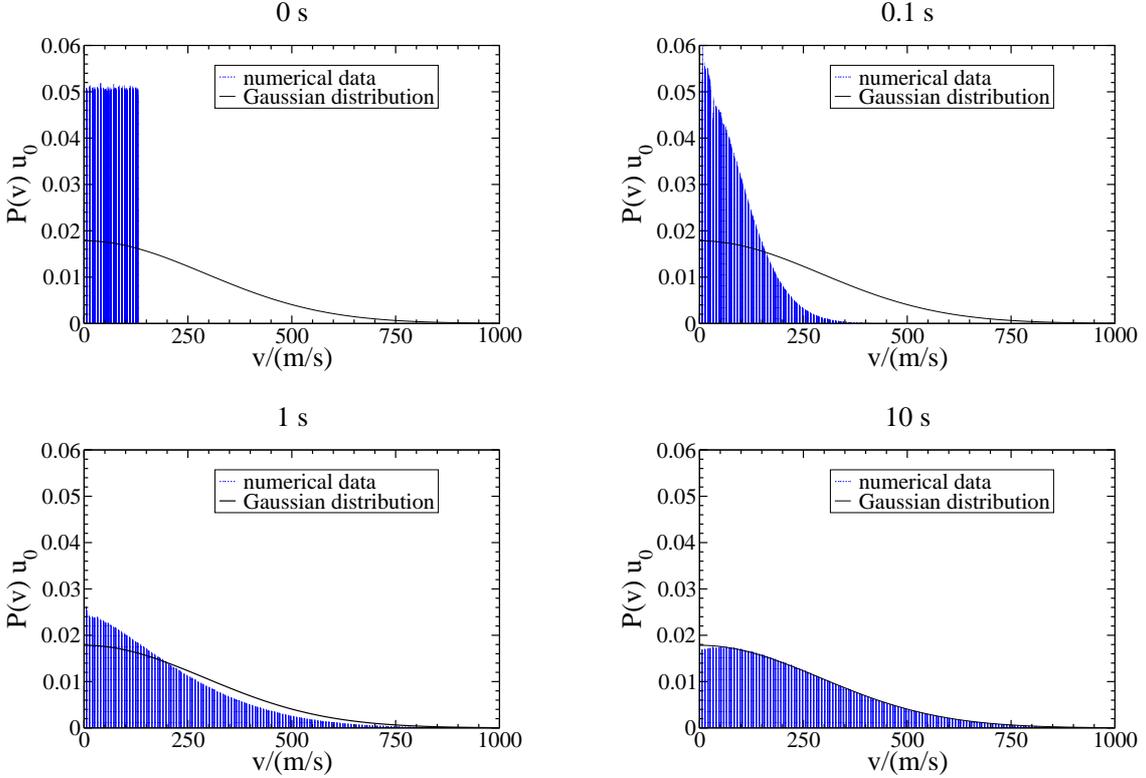

\begin{center}
\includegraphics[width=68mm]{0sec.eps}
\hfill
\includegraphics[width=68mm]{0.1sec.eps}\\[12pt]
\includegraphics[width=68mm]{1sec.eps}
\hfill
\includegraphics[width=68mm]{10sec.eps}
\end{center}
\caption{Velocity distribution $P(v>0,t)$ in the oscillator model
for $N/2=1\cdot 10^7$ particles for $v>0$
at times $t_0=0$, $t_1=0.1\, {\rm s}$, $t_2=1 {\rm s}$, $t_3=10\, {\rm s}$,
corresponding to a number of $C(t_0)=0$, $C(t_1)={\cal O}(10^1)$,
$C(t_2)={\cal O}(10^2)$, $C(t_3)={\cal O}(10^3)$ wall scatterings.
Full lines: Maxwell distribution with 
$\langle v^2\rangle=Mu_0^2/(2m)$.\label{fig3}}
\end{figure}

\subsection{Results}
\label{sect:equalphase}

\subsubsection{Random oscillator phase}

In Fig.~\ref{fig3} we show the velocity distributions $P(v>0,t)$
at times $t_0=0$, $t_1=0.1\, {\rm s}$, $t_2=1 {\rm s}$, $t_3=10\, {\rm s}$,
corresponding to a number of $C(t_0)=0$, $C(t_1)={\cal O}(10^1)$,
$C(t_2)={\cal O}(10^2)$, $C(t_3)={\cal O}(10^3)$ wall scatterings.
We see that the velocity distribution
continuously transforms from the initial step-distribution 
to the desired Maxwell distribution. The calculated
variance of the essentially stationary distribution at $t=t_3$
is $\sigma^2=502 u_0^2$, the expected
value is $\langle v^2\rangle=Mu_0^2/(2m)=500 u_0^2\equiv k_{\rm B}T/m$.
Apparently, our oscillator model 
fulfills the criterion for a physical wall because it leads to
a relaxation of the initial velocity distribution to a Gaussian.
For very small velocities, 
we observe minor deviations from the Gaussian shape.
This can be attributed to the fact that our ratio $m/M$ is finite
so that the condition $|v|>(m/M) u_0$ is violated for very small $|v|$,
cf.\ Sect.~\ref{subsec:physwall}.

We started from other non-singular velocity distributions 
with $P_0(|v|>v_0)=0$,
and also recovered the Maxwell distribution as stationary state
of our simulation.
In addition, we verified numerically 
that the Boltzmann distribution for long times~$t$ is given by
\begin{equation}
f(x,v,t\gg t_{\rm tr}) =\frac{1}{2a}\Theta\left(1-\frac{|x|}{a}\right)
f_{\rm M}(v)\; .
\end{equation}
As expected from our discussion in Sect.~\ref{sect:positionrelax},
the gas particles are equally distributed in the box, and there is no
correlation between their position and their velocity.

\subsubsection{Fixed oscillator phases}

Finally, we address the issue of whether or not our model can be 
constrained further. In our simulation results shown in Fig.~\ref{fig3}, 
the phase~$\phi_{\ell,i}$ of each wall oscillator is chosen at random.
If this phase is fixed, e.g., to $\phi_{\ell,i}=0$, 
all oscillators start with the same phase at $t=0$.
When the gas particle $g_{\ell}$ reaches the wall,
$v_{\ell,0}t_{\ell,1}=a-A$, 
the first oscillator has the contact phase
$\gamma_{\ell,1}=(t_{\ell,1}\, {\rm mod}\, T_0)$ ($T_0=2\pi/\omega_0$).
After its scattering(s) off the wall particle, the gas particle
leaves the wall region at $x_{\ell,1}$ with velocity $-v_{\ell,1}<0$, 
is reflected at the left wall,
and returns to the right wall boundary at time 
$t_{\ell,2}=(x_{\ell,1}+3a-A)/v_{\ell,1}$
with velocity $v_{\ell,1}$. The contact phase of the second oscillator
is $\gamma_{\ell,2}=(t_{\ell,2}\, {\rm mod}\, T_0)$.
In general,  if $t_{\ell,n}$ is the time at which the gas particle~$g_{\ell}$ 
reaches the boundary of the right wall for the $n$th time
$t_{\ell,n}=(x_{\ell,n-1}+3a-A)/v_{\ell,n-1}$ with velocity $v_{\ell,n-1}>0$, 
the contact phase is given by
$\gamma_{\ell,n}=(t_{\ell,n}\, {\rm mod}\, T_0)$.
The scattering off the wall can be
seen as a dynamical map ($n\geq 1$),
\begin{equation}
 \Phi: (v_n,\gamma_n) \mapsto (v_{n+1},\gamma_{n+1})
\label{eq:map}
\end{equation}
for the velocity of the gas particles and their contact phase.
The map~$\Phi$ belongs to the class of Fermi--Ulam 
accelerators~\cite{LichtenbergLieberman}.

\begin{figure}[ht]
\vspace*{9pt}
\begin{center}\includegraphics[width=9cm]{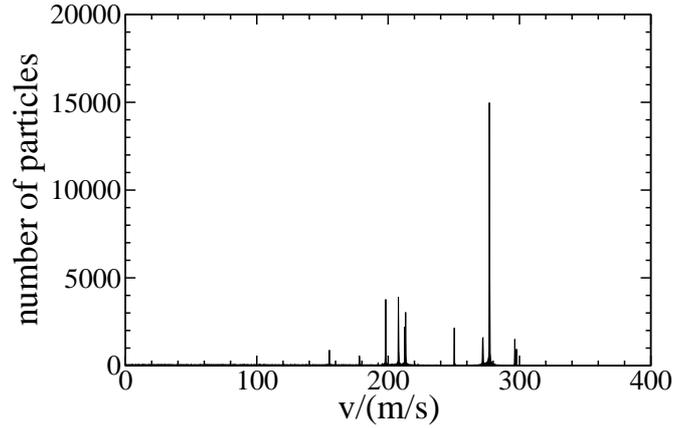}
\end{center}
\caption{Stationary velocity distribution for the gas particles 
after a large number of collisions with the dynamic wall. 
The maximal wall particle velocity
is $u_0=13\, {\rm m/s}$. The mass ratio
between gas and wall particles is $m/M=10^{-3}$,
the size of the box is $a=100 A$, and the particles evolve
according to the deterministic map~$\Phi$. 
The largest peak is a fix point of the map, 
$v_{\rm p}=v_+^*(3)=277\, {\rm m/s}$.\label{peakfig}}
\end{figure}

In general, the Fermi--Ulam accelerator map 
does not lead to a Maxwell distribution
for the gas particle velocities.
In order to see this, we note that
$\Phi$ has stationary points, $(v_{n+1},\gamma_{n+1})=
(v_{n},\gamma_{n})$. They follow from the solution
of the equations 
\begin{eqnarray}
v_n&=&\alpha v_n-\beta u(t_n)\; ,\nonumber \\
t_{n+1}-t_n=\frac{2(a+x(t_n))}{v_n}&=&\frac{2\pi R}{\omega_0} \quad 
\hbox{($R\geq 1$: integer)}\;,
\label{eq:stationarity}
\end{eqnarray}
where $u(t_n)=A\omega_0\sin(\omega_0t_n)$ 
and $x(t_n)=a-A\cos(\omega_0t_n)$.
One finds for $R>R_{\rm min}=2ra/(A\pi)\gg r$
that $v_{\pm}^*(R)=u_0[2a/A\pm
\sqrt{1-(2ra/(A\pi R))^2}]/(\pi R)$.
Moreover, the map supports stationary loops with two
and more elements. 
As we have tested numerically for $a=100A$,
some of the fix-points and loops
are attractive with a large basin of attraction around them (e.g., $R=3$).
Thus, the stationary distribution $P(v)$ depends on the initial
distribution $P_0(v)$ and consists of a few peaks only, see
Fig.~\ref{peakfig}.
In this simulation, we also observed that these peaks belong
to small values of $R$. The stationary solutions for 
large~$R$, $R\gg 1$, do not generate attractive fix-points.

The situation changes drastically for large $ra/A$, i.e. when we
choose $R_{\rm min}={\cal O}(r a/A)=10^{8}$.
As expected for a Fermi--Ulam accelerator model and confirmed
by our numerical investigations,
the map~$\Phi$ in~(\ref{eq:map}) is chaotic for $r a \gg A$;
see Refs.~\cite{Broer} for a recent introduction 
to dynamical systems and chaos.  
For example, tiny numerical errors make it impossible to retrace
the gas particles' trajectories when we reverse the time evolution 
after a large number of wall scatterings.
This implies that the deterministic map in Eq.~(\ref{eq:map}) is acceptable
for the thermalization of the gas particles in our toy model 
as long as we demand that the box is (mesoscopically)
large compared to the penetration depth
of the gas particle into the wall, $r a\gg A$.
A more detailed investigation of the map~$\Phi$
is beyond the intentions of this work.

\section{Summary and conclusions}
\label{sec:final}
In this work, we studied non-interacting point particles in a
box. 
Since we assume that the initial conditions factorize in the three
spatial dimensions, we restricted ourselves to a one-dimensional setup.
As initial condition, all gas particles start in the center of the box
with a non-singular velocity distribution. The particle positions 
`thermalize' after a large number of scatterings at the box walls, i.e., 
the gas particles become equally distributed in the box. 
This is due to the fact that
there is no conservation law for the position coordinate.
The notion of equal spatial distribution is valid
on mesoscopic length and velocity scales $\Delta x=a/\sqrt{N}$,
$\Delta v =v_{\rm typ}/\sqrt{N}$,
which requires a large number of gas particles, $N\gg 1$. 

The relaxation of the gas particles' velocity distribution 
to a Maxwell distribution is more subtle because of energy and momentum
conservation. 
We model our walls as a large collection of independent
harmonic oscillators, $L\gg N$. Each gas particle encounters
a harmonic oscillator only once (absence of a wall memory). 
The Markov property
is an essential requirement for finding a stationary velocity distribution
for the gas particles. In order to find a Maxwell velocity distribution,
we must demand in addition that there is a negligible energy transfer
between the gas and the wall particles in each collision; otherwise,
a single event could leave a trace of information of the gas particles
in the reservoir.
The condition of small energy transfers 
leads to the requirement that the wall particle
mass~$M$ is large compared to the gas particle mass~$m$, and that
their typical velocity $u_{\rm typ}$ is small compared 
with a typical gas particle velocity $v_{\rm typ}$.
Their average kinetic energies must be the same,
$M \langle u^2 \rangle =m \langle v^2\rangle$.
Finally, the oscillation amplitude of the wall particles~$A$ must be small
compared to the size of the box~$a$ so that the map of the
Fermi--Ulam accelerator guarantees a chaotic dynamics.
When these `thermodynamic' conditions are fulfilled,
the velocity distribution of the gas particles approaches a Maxwell
distribution for large times. The evolution of the full system of
gas and wall particles is fully deterministic but the result, 
a Boltzmann distribution for the gas-particle subsystem, 
is naturally obtained from statistical considerations. 
The `temperature' of the reservoir is defined by
the average kinetic energy of the wall particles.
$M \langle u^2 \rangle =k_{\rm B}T$.

Apparently, our toy model does not describe the details of the
relaxation process properly.
For example, the scattering of gas particles 
at the walls must be described quantum mechanically. As a consequence,
not every scattering process is inelastic. In fact,
most of them are elastic whereby the probability for an elastic reflection
is determined by the Debye--Waller factor. 
Therefore, the relaxation times could be (much) longer for non-interacting
gas particles than given here.
On the other hand, the interaction between gas particles and
their motion through a three-dimensional vessel reduces the relaxation time
drastically because the inter-particle scattering 
opens paths for the exchange of energy
and momentum which are not included in our toy model.
In any case, the relaxation time also depends on the shape of the
initial velocity distribution. If it is peaked around some velocities, it
takes much longer to equilibrate than for the step-like distribution used
in our example.

Despite its numerous shortcomings we hope that our work
will serve its intended main purpose, namely,
to provide a simple case study for beginners in statistical mechanics.
Our toy model illustrates that the results of measurements 
on subsystems of thermodynamically large deterministic systems 
can equally be obtained from thermodynamic considerations.

\begin{acknowledgement}
F.G.\ thanks Yixian Song for her cooperation on early stages of this project,
and S.\ Gro\ss mann, F.\ Jansson, and H.\ J\"ansch for discussions.
\end{acknowledgement}

\def\bstname{adp}
\bibliographystyle{adp}
\bibliography{relax-final}
\end{document}